\documentclass[aps,preprint,superscriptaddress]{revtex4-1}
\usepackage{graphicx}% Include figure files
\usepackage{tikz}
\usepackage{threeparttable,lipsum}
\usepackage{mathtools}
\usepackage{hyperref}
\usepackage{dcolumn}% Align table columns on decimal point
\usepackage{bm}% bold math
\DeclareRobustCommand{\VAN}[3]{#2} % set up for citation

\begin{document}

\preprint{APS/123-QED}

\title{Thermonuclear Reaction Rate of $^{30}$Si(p,$\gamma$)$^{31}$P}% Force line breaks with %\\

\author{John Dermigny}
 \email{dermigny@protonmail.com}
\author{Christian Iliadis}
 \email{iliadis@unc.edu}
\author{Art Champagne}%
\affiliation{%
	Department of Physics and Astronomy, University of North Carolina at Chapel Hill, Chapel Hill, NC 27599, USA}%
\affiliation{
	Triangle Universities Nuclear Laboratory, Durham, NC 27708, USA
}%
\author{Richard Longland}%
\affiliation{%
	Department of Physics and Astronomy, University of North Carolina at Chapel Hill, Chapel Hill, NC 27599, USA}%
\affiliation{
	Triangle Universities Nuclear Laboratory, Durham, NC 27708, USA
}%
\affiliation{
	Department of Physics, North Carolina State University, Raleigh, NC 27695, USA
}%
\date{\today}% It is always \today, today,

\begin{abstract}
Silicon synthesis in high-temperature hydrogen burning environments presents one possible avenue for the study of abundance anomalies in globular clusters. This was  suggested in a previous study, which found that the large uncertainties associated with the  $^{30}$Si(p,$\gamma$)$^{31}$P reaction rate preclude a firm understanding of the stellar conditions that give rise to the Mg-K anti-correlation observed in the globular cluster NGC 2419. In an effort to improve the reaction rate, we present new strength measurements of the $E_r^{lab} = 435$ keV and $E_r^{lab} = 501$ keV resonances in $^{30}$Si(p,$\gamma$)$^{31}$P. For the former, which was previously unobserved, we obtain a resonance strength of $\omega\gamma = (1.28 \pm 0.25$) $\times 10^{-4}$ eV. For the latter, we obtain a value of $\omega\gamma = (1.88 \pm 0.14)$ $\times 10^{-1}$ eV, which has a smaller uncertainty compared to previously measured strengths.
Based on these results, the thermonuclear reaction rate has been re-evaluated. The impact of the new measurements is to lower the reaction rate by a factor of $\approx$10 at temperatures important to the study of NGC 2419. The rate uncertainty at these temperatures has also been reduced significantly. 
\begin{description}
\item[Usage]
Secondary publications and information retrieval purposes.
\item[PACS numbers]
May be entered using the \verb+\pacs{#1}+ command.
\end{description}
\end{abstract}

\pacs{Valid PACS appear here}% PACS, the Physics and Astronomy

%\keywords{}

\maketitle

\section{Introduction}
Abundance correlations in globular clusters may provide much needed insight into the dynamical evolution of the clusters and their host galaxies. Of particular interest is NGC 2419, a globular cluster located in the outer halo of the Milky Way \cite{Mucciarelli, COKI}. A group of red giant stars in this cluster has recently been found to have an unprecedented enrichment in potassium, while simultaneously being depleted in magnesium, giving rise to a Mg-K anticorrelation among the observed stars. This observation is inexplicable within the ``single stellar population" framework commonly invoked to explain cluster evolution, and therefore hints at the existence of multiple populations. Using a simple self-pollution model, Iliadis \textit{et al.} \cite{IliadisNGC2419} and Dermigny \& Iliadis \cite{DermignyNGC} studied the stellar conditions necessary to create this puzzling signature using Monte Carlo reaction network calculations. They found that the observed abundance anomalies must have been produced at temperatures near $\approx$150~GK, for a very wide range of densities. However, the nature of the polluter stars could not be established unambiguously, in part because of large uncertainties in the thermonuclear rates of key reactions. For example, they could show that the paucity of low-energy $^{30}$Si(p,$\gamma$)$^{31}$P reaction data leads to appreciable model uncertainties, making firm conclusions difficult. Therefore, we present thermonuclear rates for the $^{30}$Si(p,$\gamma$)$^{31}$P reaction (Q = $7296.55 \pm 0.02$ keV \cite{Wang}) based on new resonance strength measurements.

\begin{figure}
%0.48
\includegraphics[width=0.6\textwidth]{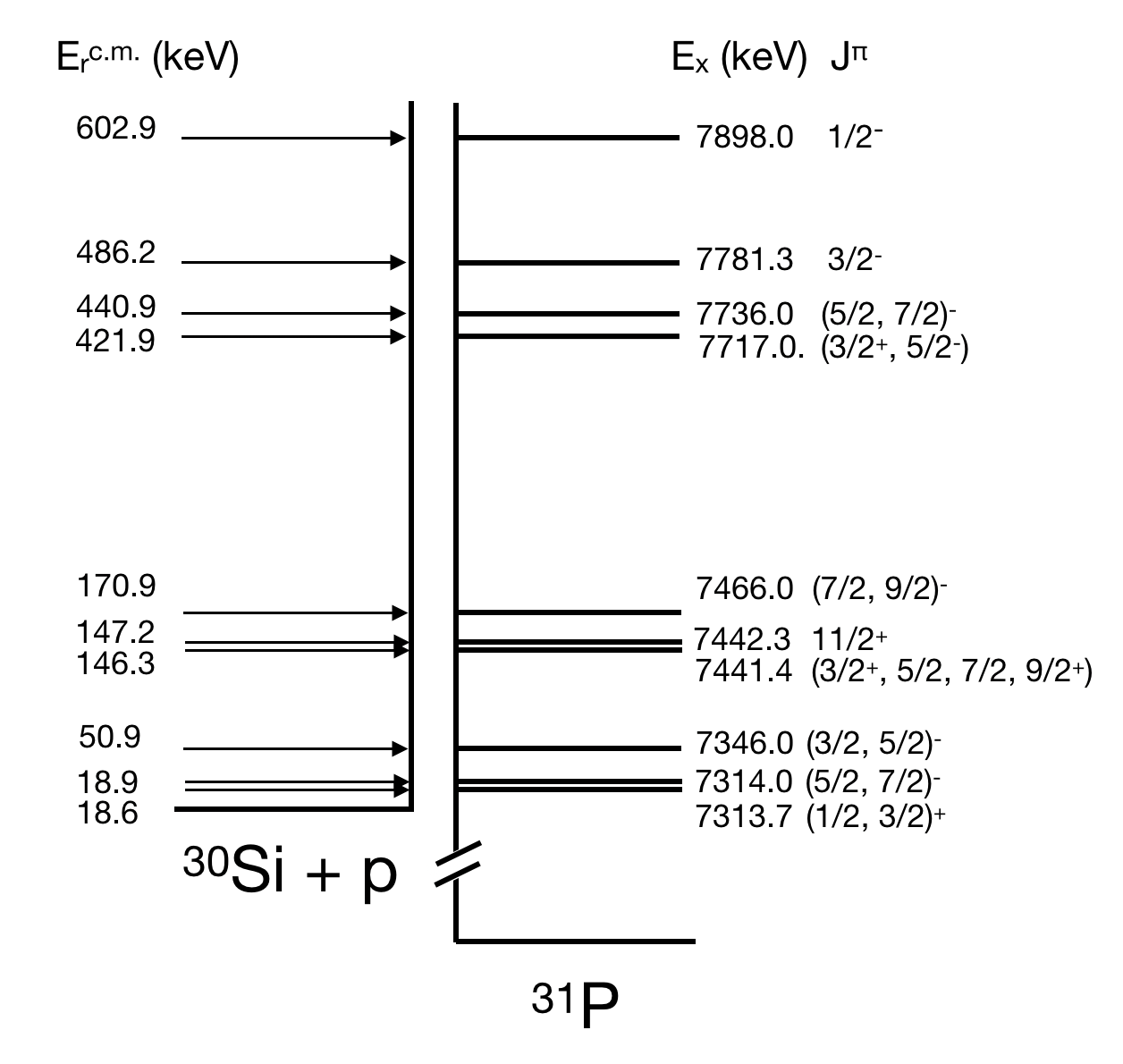}
\caption{Energy-level diagram of $^{31}$P. Only levels relevant for the present work are displayed. Excitation energies and $J^\pi$ values are adopted from multiple sources including the present work. See text for details. }
\label{fig:level_scheme}
\end{figure}

The level structure of $^{31}$P near the proton threshold is shown in Figure~\ref{fig:level_scheme}. The first modern reaction rate for  $^{30}$Si(p,$\gamma$)$^{31}$P, based on experimentally derived data, was published by Harris, Caughlan \& Fowler \cite{ReactionRatesIII}. However, no reaction rate uncertainty was provided. The rate was re-evaluated by Iliadis \textit{et al.} \cite{Iliadis2001}. Their work featured two major improvements over that of Ref.~\cite{ReactionRatesIII}. First, the effects of the unobserved threshold states between E$_{r}^{c.m.} = 52$ keV $-$ $416$ keV were included in the calculation. Second, statistical and systematic uncertainties in the experimental data were propagated through to the final reaction rate, affording an estimation of the temperature region where the rates are most uncertain. Most recently, the rates were evaluated again by Iliadis \textit{et al.} \cite{ILIADIS2010b} using the Monte Carlo reaction rate formalism by Longland \textit{et al.} \cite{LONGLAND2010}. This rate was used in the nucleosynthesis studies by Iliadis \textit{et al.} \cite{IliadisNGC2419} and Dermigny \& Iliadis  \cite{DermignyNGC}. 

The two latter works explored stellar temperatures between approximately 160 MK and 300 MK as a means of explaining the abundance anomalies observed in NGC 2419. At these temperatures, the $^{30}$Si(p,$\gamma$)$^{31}$P reaction rate is determined by the resonances at $E_r^{lab}=435$ keV ($E_{cm}^{lab}=422$ keV) and $E_r^{lab}=501$ keV ($E_{cm}^{lab}=486$ keV). Little is known about the $435$-keV resonance. The corresponding $^{31}$P compound nucleus level was populated by Vernotte \textit{et al.} \cite{VERNOTTE1990} using the $^{30}$Si($^{3}$He,d)$^{31}$P proton transfer reaction, but the resonance has not yet been measured directly. Consequently, the most recent rate evaluation had to rely on the upper-limit formalism of Ref.~\cite{LONGLAND2010}. An additional complication is introduced by the unknown spin-parity assignment of this level.
\par
The $501$-keV resonance has been measured previously by Hoogenboom \textit{et al.}~\cite{Hoogenboom} and Riihonen \textit{et al.} \cite{RIIHONEN1979}. However, their reported resonance strengths differ by approximately a factor of two. The uncertainties associated with both of these resonances give rise to a factor of three uncertainty in the reaction rate in the important stellar temperature range. 
\par
In the present work, we report on new measurements of the resonance strengths for the 435-keV and 501-keV resonances in $^{30}$Si(p,$\gamma$)$^{31}$P. 
These measurements are critical to achieving the  improved accuracy required for modern nucleosynthesis calculations and are of particular interest to the observations in NGC 2419. See Dermigny \& Iliadis \cite{DermignyNGC} for further detail.

%These were performed using a thick-target yield curve analysis, where a proton beam was incident on a implanted silicon target. The analysis of the pulse-height spectra was performed using the fraction fitting methodology presented in \cite{DERMIGNY2016}. In addition to the resonance strengths, the $\gamma$-ray decay paths and probabilities from the two compound states to the ground state of $^{31}$P were studied and quantified.

\par
In Section \ref{Equipment} we describe the experimental apparatus. In Section III we discuss the data analysis techniques. The results of the experiment are then given in Section IV.  New thermonuclear rates for $^{30}$Si(p,$\gamma$)$^{31}$P are derived in Section V. A summary and conclusion is given in Section VI.
\section{Equipment} \label{Equipment}
\subsection{Accelerators and targets} \label{targets}
The experiment was carried out at the Laboratory for Experimental Nuclear Astrophysics (LENA), which is part of the Triangle Universities Nuclear Laboratory (TUNL) \cite{CESARATTO2010}.
LENA houses a JN Van de Graaff electrostatic accelerator. 
The JN Van de Graaff is capable of delivering up to 150 $\mu$A of protons to target at energies between 200 keV and 900 keV \footnote{The JN Van de Graaff accelerator is being decommissioned at the time of writing and will be replaced by a new machine.}. 
Prior to the resonance experiments, the proton beam-energy was calibrated via a yield curve analysis of several well-known resonances in the $^{27}$Al(p,$\gamma$)$^{28}$Si reaction. Several measurements were also made using the $^{12}$C(p,$\gamma$)$^{13}$N direct-capture reaction. 
Using these two complementary techiques, a beam-energy uncertainty of $\approx$ $1$ keV was determined. The full width at half maximum of the beam-energy profile observed during these experiments was $0.8$ keV.

\par
The proton beam from the JN accelerator entered the target chamber through a liquid-nitrogen-cooled copper tube. An electrode was mounted at the end of this tube and was biased to $-300$ V to suppress the emission of secondary electrons from the target and the beam collimator. The target and chamber formed a Faraday cup for charge integration. The beam was focused and rastered into a circular profile of $\approx$ $9$ mm diameter on target. The target was directly water cooled using deionized water.
\par
 \begin{figure}
 %0.5
	\includegraphics[width=0.6\textwidth]{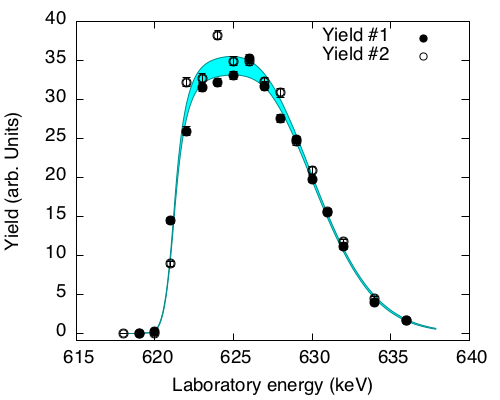}
	\caption{Yield curves of the E$_r^{lab}=622$ keV resonance in $^{30}$Si(p,$\gamma$)$^{31}$P. Both yields were measured at the start of the experiment in succession.  The uncertainties shown derive from counting statistics only. The cyan area represents the $95\%$ credible region, as determined using a Bayesian method, to extract the maximum yield, target thickness, and area under the yield curve.}
	\label{fig:yield}
\end{figure}

The target was fabricated at TUNL by implanting $^{30}$Si ions into a 0.5-mm-thick tantalum backing. A 55-keV $^{30}$Si beam was generated from isotopically enriched silicon powder ($99.64\%$) by a SNICS source \cite{MIDDLETON1984}.
The total dose incident on the backing was 360 mC. Prior to implantation, the tantalum backings were chemically etched and then outgassed in high vacuum by resistive heating to remove contaminants.
The well-known resonance at $E_r^{lab}=622$ keV \cite{ENDT1990} in $^{30}$Si(p,$\gamma$)$^{31}$P was used to characterize the target. Yield curves are shown in Figure~\ref{fig:yield}. The target thickness was found to be  $8.7\pm0.1$ keV. Based on the maximum yield obtained and the resonance strength reported in Ref.~\cite{PAINE1979}, the stoichiometry of the target layer was Ta:$^{30}$Si = $1:1.50 \pm 0.24$. %Ta$_{5}$$^{30}$Si$_{7.5\pm1.2}$. 
Yield curves measured at the end of the experiment demonstrated that the maximum yield and thickness were unchanged after an accumulated proton charge of 5 C.
\subsection{Spectrometer}
\begin{figure}
 %0.5
 	\includegraphics[width=0.6\textwidth]{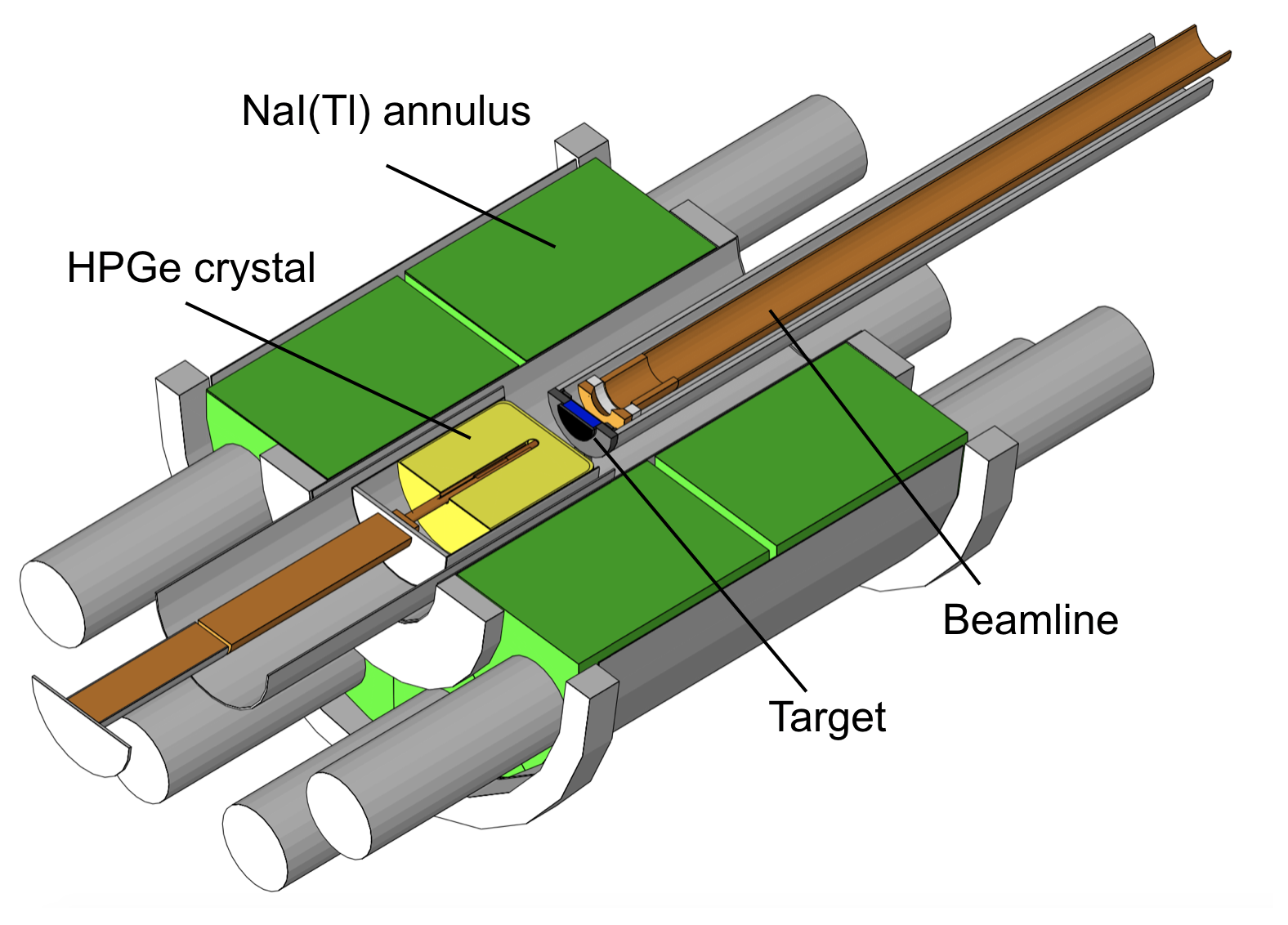}
 	\caption{The LENA $\gamma\gamma$-coincidence spectrometer. A $135\%$ HPGe detector (yellow) is surrounded by a 16-segment NaI(Tl) annulus (green). The HPGe detector is located in close proximity to the target chamber for maximum efficiency. Not shown are the plastic scintillator paddles.}
 	\label{fig:detector}
 \end{figure}

 The $^{30}$Si($p$,$\gamma$)$^{31}$P reaction gives rise to the emission of multiple, coincident $\gamma$-rays. The $\gamma\gamma$-coincidence spectrometer employed at LENA has been designed to exploit this property in order to improve detection sensitivity \cite{Longland:2006tz}. 
 The setup is shown in Figure \ref{fig:detector}. It features a $135\%$ HPGe detector, oriented at $0^{\circ}$ with respect to the beam-axis, placed in close proximity to the target chamber. The HPGe detector is surrounded by a 16-segment NaI(Tl) annulus. Both counters are surrounded on five sides by 50-mm-thick plastic scintillator panels (not shown in the figure) to suppress cosmic-ray muon events.
 \par
 %The HPGe signals serve as the master trigger for the data acquisition. 
 Energy and timing information from each detector was processed using standard NIM and VME electronics. Events were sorted off-line using the acquisition software JAM \cite{JAM}. Coincidence conditions were then applied in software by constructing a two-dimensional NaI(Tl) versus HPGe detector energy spectrum and applying a trapezoidal gate with the following condition:
 \begin{equation}
 3.5 \text{ MeV} < E_{\text{HPGe}} + E_{\text{NaI(Tl)}} < 9.0 \text{ MeV. }
 \end{equation}
 If an event satisfied this condition, the HPGe detector signal was included in the gated pulse-height spectrum. The upper threshold of $9.0$ MeV was chosen to exclude cosmic-ray induced background events with an energy exceeding the $^{31}$P excitation energy range of interest. The low-energy threshold of $3.5$ MeV significantly reduces environmental background (e.g., $^{40}$K, $^{208}$Tl), as well as beam-induced  background from contaminant reactions with relatively small Q-values, e.g., $^{12}$C+p (Q = $1943.49 \pm 0.27$ keV \cite{Wang}).
 \par
The spectrometer detector dead time was monitored throughout the experiment by feeding a pulser signal into the HPGe preamp. The number of pulses was recorded using a scalar counting module, and this was compared to the artificial pulser peak in the pulse-height spectrum to obtain the dead time of the system. The dead time was kept below 5\% to avoid pulse pileup effects.
\par
 
\section{Data Analysis Strategy}
The extraction of accurate resonance strengths from singles and coincidence data requires careful calibrations and corrections for experimental artifacts, e.g., detection efficiencies, coincidence summing, angular correlations, and finite beam spot sizes. The most reliable way to correct for these effects is to employ Monte Carlo radiation transport codes, such as Geant4 \cite{GEANT4}. This requires that the detector geometry is precisely known \cite{Carson:2010kh}. A complete Geant4 model of our spectrometer, including the beam tube, target holder, and passive shielding, is presented in Howard \textit{et al.} \cite{HOWARD2013}. Furthermore, we  recently developed a method to analyze not just the net intensity of isolated peaks in the pulse height spectra, but to fit large energy regions of measured singles and gated HPGe detector spectra using a binned-likelihood approach based on a Bayesian method \cite{DERMIGNY2016}. This technique has already been successfully applied to the analysis of $^{17}$O(p,$\gamma$)$^{18}$F \cite{BUCKNER_2015} and $^{22}$Ne(p,$\gamma$)$^{23}$Na \cite{KELLY2017} reaction data.
\par
The main feature of this method is that the spectra are analyzed in a bin-by-bin fashion, where the fit is based on the simulated response of the detector system to all sources of radiation present during the experiment. The advantage over traditional analysis methods is that the simulated response implicitly includes the effects mentioned above, removing the need for cumbersome individual corrections.
\par
The procedure for building a successful model is straightforward. For each experimental singles or coincidence spectrum, we prepared several component spectra (``templates''). For example, a spectrum measured on the plateau of a thick-target yield curve typically has contributions from several primary decays and a number of beam-induced and environmental background components. The response of the $\gamma\gamma$-spectrometer to each individual primary decay was simulated using Geant4, incorporating all subsequent secondary transitions. The required secondary $\gamma$-ray branching ratios were adopted from Ref. \cite{de_Neijs_1975}.
Angular correlation effects for the decaying states were also included in the simulations to the extent possible. If they have been measured previously for any of the primary decays, those results were used to simulate the $\gamma$-ray emission pattern. In cases where the spin-parity of the decaying states permitted the exact calculation of the angular correlation, those were incorporated into the simulation instead.
 A measured spectrum, obtained without beam on target, served as the room-background template. For the beam-induced background components, templates were generated using a Geant4 simulation of the contaminant reactions.  
\par
Once the templates have been generated, the experimental data were fit using a Bayesian binned-likelihood probability model \cite{barlow}. The fit provides the scaling parameter of each template. The scaling parameters are then used to determine the total number of $^{30}$Si(p,$\gamma$)$^{31}$P reactions, including the primary branching ratios for $^{30}$Si(p,$\gamma$)$^{31}$P. 
\par 
\par
In previous applications of fraction-fitting, for example, in Buckner \textit{et al.} \cite{BUCKNER_2015}, Kelly \textit{et al.} \cite{KELLY2017}, and Dermigny \textit{et al.} \cite{DERMIGNY2016}, the experimental spectra were fit using a single, contiguous region spanning several MeV. In this work, a different approach was taken. Instead, the fits were limited to regions surrounding the primary transition full-energy peaks. Bins that fell outside of these boundaries did not enter into the resonance strength calculations. These regions were approximately 50-keV wide, with the primary peak located in the center. This was done in order to minimize the influence of the secondary transition $\gamma$-ray branching ratios that were adopted from the literature, since the primary transition $\gamma$-ray intensities are the most important to the resonance strength calculation.
%The change in methodology was necessitated by the difficulties faced in fitting the observed secondary transition full-energy peaks, which were frequently over- or under-predicted in our simulated template spectra. 
\par
\section{Results}
\subsection{Resonance at \texorpdfstring{$E_{r}^{\text{lab}} = 622$}{Er(lab) = 622} keV}
The $E_r^{lab}=622$ keV resonance has been measured several times \cite{SMITH1958,HOUGH1968,LYONS1969,RIIHONEN1979}, but the early measurements were in mutual disagreement. For this reason, Paine \& Sargood \cite{PAINE1979} remeasured the resonance as part of a campaign to improve resonance strengths in the Z=11-20 region. They reported on several relative and absolute measurements of the 622-keV resonance strength. The precision and consistency of their work helped to make it a standard resonance \cite{Iliadis2001} for measuring the lower-energy resonances in the $^{30}$Si(p,$\gamma$)$^{31}$P reaction. For this purpose, we adopted their recommended value, $\omega\gamma_{622} = 1.95 \pm 0.10$ eV.
\par
The $E_r^{lab}=622$-keV resonance corresponds to a $^{31}$P compound level at $E_x=7896 \pm 1$ keV \cite{OUELLET2013}. The $\gamma$-ray decay of this level is well-known, and the branching ratios are presented in Ref.~\cite{de_Neijs_1975}. The spin-parity has been determined using proton transfer studies and $\gamma\gamma$ angular correlation measurements, which support an assignment of $J^{\pi}= 1/2^-$ \cite{Broude_1958, VERNOTTE1990}. Since our incident proton beam is unpolarized, conservation of angular momentum dictates that the primary transition $\gamma$-rays are emitted isotropically. This further simplified the analysis by eliminating the need for angular correlation effects in our simulations.
\par
The measured yield curve for the $E_{p}^{lab} = 622$ keV resonance is shown in Figure~\ref{fig:yield}. The net intensity of the ground-state transition $\gamma$-ray in the singles spectrum was used to calculate the yield.
To obtain high-statistics resonance data, a longer run was performed on the plateau maximum at an incident proton energy of $E_p^{lab}=625$ keV. The total amount of charge accumulated for this measurement was 3781 $\mu$C, with an average beam-intensity of 0.79 $\mu$A on target.
\par
We then fit the data using the fraction-fitting method. For the singles and coincidence spectra, the derived primary branching ratios as well as the total number of $^{30}$Si(p,$\gamma$)$^{31}$P reactions are shown in agreement in Table~\ref{fig:results}. All transitions from the compound state that had been previously identified by de Neijs \textit{et al}. \cite{de_Neijs_1975} were present in the acquired spectra. No new transitions were observed. The branching ratios measured are consistent with those reported by de Neijs \textit{et al.} \cite{de_Neijs_1975}, although a quantitative comparison is difficult since the latter work did not report uncertainties for their branching ratios.
\par
From the measured primary $\gamma$-ray energies, we find an excitation energy of $E_x$ $=$ $7898.0\pm0.3$~keV. Our result agrees with the previously reported value but has a smaller uncertainty \cite{OUELLET2013}. This excitation energy corresponds to a laboratory and center-of-mass resonance energy of $E_r^{lab}$ $=$ $621.7\pm0.3$~keV and $E_r^{cm}$ $=$ $602.9\pm0.3$~keV, respectively (see Table I).

\linespread{1.0}
\begin{table*}[ht!] 
\begin{threeparttable}
\caption{Total number of $^{30}$Si(p,$\gamma$)$^{31}$P reactions and primary branching ratios for low-energy resonances}

%%%Measurements of the reaction intensities and primary branching ratios for the E$_p^{lab}$ = $435$ keV, $498$ keV, and $620$ keV resonances in the $^{30}$Si(p,$\gamma$)$^{31}$P reaction. The present results were obtained from the singles and coincidence spectra using the fraction-fitting method. The branching ratios obtained by De Neijs \textit{et al.} \cite{de_Neijs_1975} are shown in the last column.
        % Appropriate for two-columns mode
		%\setlength{\tabcolsep}{7pt}
		% Appropriate for one-column mode
		\setlength{\tabcolsep}{1pt}

		\center
		\begin{tabular}{@{}lccccllccc@{}}
		\toprule \\
	 	$E_r^{lab}$ (keV)\tnote{a} & $E_x$ (keV)\tnote{a} & $J^{\pi}$ &\multicolumn{2}{c}{Number of $^{30}$Si+p Reactions\tnote{a}}& & \multicolumn{4}{c}{Branching Ratios ($\%$)}\\
		\colrule \\
	    & & & singles  & coincidence & 		& Transition & singles\tnote{a}  & coincidence\tnote{a} & Ref. \cite{de_Neijs_1975} \\
	      \\
	621.7$\pm$0.3 & 7898.0$\pm$0.3 & $\frac{1}{2}^-$\tnote{b} & $15.10(7) \times 10^{6}$ & $14.69(10) \times 10^{6}$ && R $\rightarrow$ 0      & $94.4(2)$       &  $94.5(2)$    &  $95$    \\	
	    & & & &  && R $\rightarrow$ 1266                     &  $1.79(13)$     &  $1.85(18)$   &  $1.4$   \\		
	    & & & &  && R $\rightarrow$ 3134   &  $0.65(6)$      &   $0.53(7)$     &  $0.6$    \\		
	    & & & &  && R $\rightarrow$ 3506   &   $0.53(5)$     &  $0.47(6)$      & $0.5$         \\		
	    & & & &  && R $\rightarrow$ 5015   &  $2.63(6)$      &   $2.63(7)$     &  $2.5$     \\		
  \\
 	501.1$\pm$0.2 & 7781.3$\pm$0.2 & $\frac{3}{2}^-$\tnote{b} & $1.51(7) \times 10^{7}$ & $1.47(6) \times 10^{7}$ &&		R $\rightarrow$ 0   &  $50.2(4)$    &  $47.5(2)$  & $52$ \\	
 	
  & & & &  &&		R $\rightarrow$ 1266         &  $26.8(3)$    &  $28.3(2)$  & $27$ \\

&& && && R $\rightarrow$ 2233  	    &         $4.8(2)$   &  $4.7(2)$&  $5.0$  \\		
&& && && R $\rightarrow$ 3134      &        $10.4(2)$   & $10.6(2)$ & $11$   \\		
&& && && R $\rightarrow$ 3295      &         $0.8(1)$  &  $0.6(1)$  &  $0.6$ \\	
&& && && R $\rightarrow$ 4260      &         $0.6(1)$  &  $0.6(1)$  &  $0.5$ \\		
&& && && R $\rightarrow$ 4783      &         $2.7(1)$  &  $2.7(1)$ & $2.3$   \\		
&& && && R $\rightarrow$ 5014.9    &         $2.4(1)$  &  $3.4(1)$ & $1.6$   \\		
&& && && R $\rightarrow$ 5116      &         $0.5(1)$  &  $0.5(1)$  & ---   \\	
&& && && R $\rightarrow$ 6496      &         $0.5(1)$  &  $0.5(1)$&  --- \\	 					
&& && && R $\rightarrow$ 6594      &         $0.4(1)$  &  $0.4(1)$ &  ---  \\
  \\
434.6$\pm$0.3 & 7717.0$\pm$0.3 & ($\frac{3}{2}^+$, $\frac{5}{2}^-$)\tnote{c} & $5.4(1) \times 10^{5}$ & $5.3(1) \times 10^{5}$ & & R $\rightarrow$ 3295   &   $23.6(10)$     &   $20.7(10)$   &  ---     \\		
&& && && R $\rightarrow$ 4431       &    $38.5(10)$    &  $41.5(10)$    &  ---       \\		
&& && && R $\rightarrow$ 5014.9     &   $37.9(9)$      &   $37.8(10)$   &  ---       \\	
\\
  \botrule \\

\end{tabular}
\begin{tablenotes}
\item[a] {\footnotesize Present work. The Number of $^{30}$Si+p reactions was obtained from singles and coincidence spectra using the fraction-fitting method. For the $435$~keV resonance, the values listed in columns 4, 5, 7, and 8 do not include corrections for angular correlation results. These were applied {\it ex post} (see text).}
\item[b] {\footnotesize From Ref. \cite{OUELLET2013}.}
%\item[c] {\footnotesize Present work.}
\item[c] {\footnotesize Including information from the present work (see text).}
\end{tablenotes}
\label{fig:results}
\end{threeparttable}
\end{table*}    
\linespread{1.5}

%%%%%%%%%%%%%%%%%%%%%%%%%%%%%%%%%%%%%%%%%%%%%%%%%%%%%%%%
\subsection{Resonance at $E_{r}^{\text{lab}} = 501$ keV}
The strength of the resonance at $E_{p}^{lab}=501$ keV ($J^\pi=3/2^-$ \cite{Broude_B,RIIHONEN1979}) has been measured previously by Hoogenboom \textit{et al.} ($\omega\gamma=0.086\pm0.008$ eV, unpublished thesis, see Endt \cite{ENDT_1967}) and Riihonen \textit{et al.} ($\omega\gamma=0.165\pm0.025$ eV \cite{RIIHONEN1979}). These values are in conflict, differing by nearly a factor of two.
Unfortunately, we can only speculate as to the cause of this discrepancy since virtually no information from the Hoogenboom \textit{et al.} measurement or analysis is available today. With that in mind, it seems their value has had an out-sized impact on previous $^{30}$Si(p,$\gamma$)$^{31}$P reaction rate calculations. For instance, the Iliadis \textit{et al.}~\cite{ILIADIS2010b} evaluation adopted the weighted-average for the $E_{p}^{lab}=501$ keV resonance strength suggested by Ref.~\cite{ENDT1990}. Because of the small uncertainty reported by Hoogenboom \textit{et al.}, neither the resonance strength nor the calculated reaction rate reflected the tension underlying these two measurements.

To improve this situation we remeasured the resonance strength.
We first obtained a yield curve over the $E_{p}^{lab}=501$ keV resonance using the net-intensity of the R$\rightarrow$1266 transition. At $E_p^{lab}=503$ keV, corresponding to the plateau maximum,  high-statistics resonance data were recorded. The total incident charge was $40$ mC, with an average beam intensity of $13$ $\mu$A on target. In addition to the seven transitions reported by de Neijs \textit{et al.}~\cite{de_Neijs_1975}, we identified three full-energy peaks in the pulse-height spectra that correspond to previously unobserved transitions. Based on their energies, we have identified them as primary transitions from the $^{31}$P compound state to the excited states at E$_x=5116$ keV, E$_x=6496$ keV, and E$_x=6594$ keV. The three full-energy peaks were found in a region of the pulse-height spectrum encumbered by both environmental and beam-induced background, and so it is likely that they had escaped detection in Ref.~\cite{de_Neijs_1975}.

The measured singles and coincidence data were then analyzed via fraction-fitting. 
The resulting reaction intensities and branching ratios are shown in Table~\ref{fig:results}. In general there is good agreement between the branching ratios obtained using the singles and the coincidence data, with the exception of the R$\rightarrow5014.9$ primary transition. This state is part of a doublet ($E_x=5014.9$ keV and $E_x=5015.0$ keV) and so it is conceivable that the secondary branching ratios reported in Ref.~\cite{de_Neijs_1975} contain errors. Since this information is used to generate our templates, such an error would manifest itself in the coincidence results.
With regard to the primary transition branching ratio values reported by de Neijs \textit{et al.} \cite{de_Neijs_1975}, the lack of uncertainties again makes a quantitative comparison difficult. This is further complicated by the present observation of the three new transitions.
\par
From the measured primary $\gamma$-ray energies, we find an excitation energy of $E_x$ $=$ $7781.3\pm0.2$~keV. Our result has a smaller uncertainty than the previously reported value of $7779\pm1$~keV but disagrees at the one-sigma level \cite{OUELLET2013}. 
This excitation energy corresponds to a laboratory and center-of-mass resonance energy of $E_r^{lab}$ $=$ $501.1\pm0.2$~keV and $E_r^{cm}$ $=$ $486.2\pm0.2$~keV, respectively (see Table I).
\par
The resonance strength was calculated using the relative measurement formula \cite{Iliadis_2015}:
\begin{equation}
\frac{\omega\gamma_{501}}{\omega\gamma_{622}} = \frac{\lambda^2_{622}}{\lambda^2_{501}} \frac{(B\eta W)_{622}}{(B \eta W)_{501}} \frac{A_{Y,501}}{A_{Y,622}} \text{,}
\end{equation}
where $\omega\gamma_{622}$ is the resonance strength of our standard resonance, $\lambda_i$ is the de Broglie wavelength of the incident proton, and $A_{Y,i}$ is the area under the yield curve for resonance $i$. The factors $B$, $\eta$, and $W$, refer to the branching ratio, detection efficiency, and angular correlation coefficient, respectively, for the observed primary transition. The $A_{Y,i}$ values were obtained using a fit to the resonance yield curves. The combined correction factor $(B \eta W)_i$ was calculated for each resonance $i$ and the respective transition, R$\rightarrow$ E$_x$, using the ratio:
\begin{equation}
    (B \eta W)_i = \frac{I_{R \rightarrow E_x}}{N_{total,i}},
\end{equation} 
where $N_{total}$ is the total number of $^{30}$Si $+$ $p$  reactions and $I_x$ is the net-intensity of the primary transition full-energy peak. The latter quantity was measured directly from the singles pulse-height spectrum.
\par
We determined a resonance strength of $\omega\gamma_{501}=0.188 \pm 0.014$ eV (Table~\ref{tab:levels}). This is in conflict with the original measurement by Hoogenboom \textit{et al.}, while being consistent with the more recent Riihonen \textit{et al.} value. The effect of this new measurement on the reaction rate will be considered in Section V.
\par
%%%%%%%%%%%%%%%%%%%%%%%%%%%%%%%%%%%%%%%%%%%%%%%%%%%%%%%%%%%
\subsection{Resonance at $E_{r}^{\text{lab}} = 435$ keV}
Unlike the $622$-keV and $501$-keV resonances, for which measured spin-parity and deexcitation branching ratios are available, little is known about the resonance at E$_r^{lab}=435$ keV. Previous measurements of this state are limited to a few indirect studies that were unable to determine either a spin-parity or a single-particle reduced width \cite{Betigeri,MOSS1969,AlJadir,VERNOTTE1990}. 
Estimates of the resonance strength and its effect on the $^{30}$Si(p,$\gamma$)$^{31}$P reaction rate have therefore been limited to experimental upper-limits and statistical arguments (see Iliadis \textit{et al.} \cite{ILIADIS2010b}). 

\begin{figure*}[ht!]
\center
 	\includegraphics[width=0.95\textwidth]{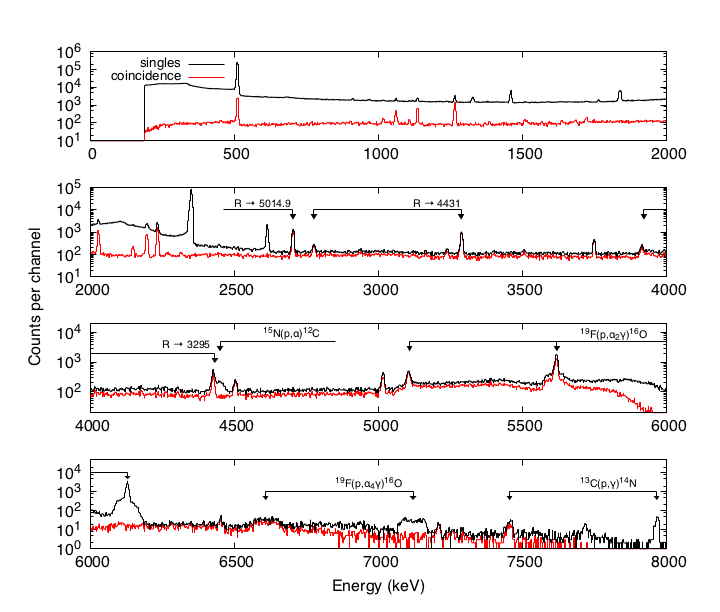}
 	\caption{Measured singles (black) and coincidence (red) pulse-height spectra collected at the 435-keV resonance. Full-energy peaks are indicated for the primary transitions in $^{30}$Si(p,$\gamma$)$^{31}$P and beam-induced contaminant reactions.}
 	\label{fig:435-spectra}
 \end{figure*}

We obtained the first $\gamma$-ray spectra of the $435$-keV resonance using a proton beam with $E_p^{lab}=437$ keV, corresponding to the maximum of the yield curve. The average beam current on target was about $80$ $\mu$A, with a total accumulated charge of $2$ C. The pulse-height spectra for the singles and coincidence data are shown in Figure~\ref{fig:435-spectra}.
Three primary transition full-energy peaks are indicated. Based on their energies, we have identified them as transitions from the $7718$ keV compound state to the states at E$_x=3295$ keV ($J^\pi$ $=$ $5/2^+$ \cite{OUELLET2013}), E$_x=4431$ keV ($J^\pi$ $=$ $7/2^-$ \cite{OUELLET2013}), and E$_x=5014.9$ keV ($J^\pi$ $=$ ($3/2^+$) \cite{OUELLET2013}). The last assignment was established based on the presence of the $5014.9\rightarrow0$ and $5014.9\rightarrow1266$ secondary transitions. The relative intensity of these two full-energy peaks was found to be consistent with the branching ratios for the decay of the $5014.9$-keV state published by de Neijs \textit{et al.} \cite{de_Neijs_1975}. 
\par
 
 From the measured primary $\gamma$-ray energies, we find an excitation energy of $E_x$ $=$ $7717.0\pm0.3$~keV. Our result agrees with the previously reported value of $7718\pm4$~keV but has a smaller uncertainty \cite{OUELLET2013}. This excitation energy corresponds to a laboratory and center-of-mass resonance energy of $E_r^{lab}$ $=$ $434.6\pm0.3$~keV and $E_r^{cm}$ $=$ $421.9\pm0.3$~keV, respectively (see Table I).
\par
The spin-parity is not unambiguously known for this level, but can be restricted based on the $\gamma$-ray decay observed in the present work (Table~I) and previous transfer-reaction measurements \cite{AlJadir}. The primary decays proceed to  levels with unambiguous assignments of $J^\pi$ $=$ $5/2^+$ and $7/2^-$. Using Endt's ``Dipole or E2 rule'' \cite{ENDT1990}, we find $J^\pi$ $=$ ($3/2^+$, $5/2$, $7/2$, $9/2^+$). The $^{29}$Si($^3$He,p)$^{31}$P transfer-reaction measurement by Al-Jadir \textit{et al.}~\cite{AlJadir} suggests spin-parity values of $J^\pi$ $=$ ($1/2^+$,$3/2$, $5/2^-$). 
Combining these observations we find $J^\pi$ $=$ ($3/2^+$, $5/2^-$) for the $435$-keV resonance. 

Angular correlation corrections were applied as follows. We introduced the factor $\xi$, given by:
\begin{equation}
    \xi = \frac{  \langle N_{\text{total}}^{corr}\rangle  }{N_{\text{total}}^{meas}}\;,
\end{equation}
where $N_{\text{total}}^{meas}$ is the total reaction intensity obtained using the approximation of isotropy and $N_{\text{total}}^{corr}$ is the estimated true reaction intensity. This latter result was obtained using a Monte Carlo procedure. For each iteration, the spin-parity of the resonance was sampled from a set of plausible values. The sampled spin-parity is then used to calculate the possible decay channels for each transition (See Appendix D in Iliadis 2015 \cite{Iliadis_2015} for details). For the estimation of $N_{\text{total}}^{corr}$, we only considered contributions from E1, M1/E2, and E2 radiation. If a transition had a mixed radiation decay, we sampled the mixing ratio from a uniform distribution over the interval $[0,1]$. Each iteration of our procedure then yielded a value of $N_{\text{total}}^{corr}$ given by:
\begin{equation}
  N_{\text{total}}^{corr}= \sum_{j}{ \frac{N^{R \rightarrow j}}{W^j}}\; ,
\end{equation}
where for each primary transition $j$, the angular correlation $W^j$ has been calculated based on randomly sampled reaction parameters and is used to correct the measured transition intensity, $N^{R\rightarrow j}$. 
The $50$th percentile of the distribution for $N_{\text{total}}^{corr}$  is then used to define $\langle N_{\text{total}}^{corr} \rangle$ and the uncertainty is given by the $16$th and $84$th percentiles. Using this procedure, we determined a correction factor of $\xi=1.18\pm0.20$. 
The large uncertainty reflects the wide range of angular correlation effects possible within the narrow set of spin-parities values.
\par
Finally, we calculated the resonance strength using the thick-target relative measurement formula \cite{Iliadis_2015}:
\begin{equation}
    \frac{\omega\gamma_{435}}{\omega\gamma_{622}} = \frac{\epsilon_{\text{eff},435}}{\epsilon_{\text{eff},622}} \frac{\lambda_{622}^2}{\lambda_{435}^2} \frac{Y_{max,435}}{Y_{max,622}},
\end{equation}
where $Y_{max,i}$ is the maximum yield and $\epsilon_{\text{eff},i}$ is the effective stopping power at resonance energy $i$.
The maximum yield is given by the ratio of the reaction intensity, $N_{\text{total}}$, to the number of incident protons, $N_p$, including the correction factor $\xi$:
\begin{equation}
Y_{max,435} = \xi N_{\text{total},435}/N_{p,435} .
\end{equation}
The effective stopping powers were calculated at the resonance energies using the relationship \cite{Iliadis_2015}:
\begin{equation}
    \epsilon_{\text{eff},i} = \epsilon_{Si} + \frac{N_{Ta}}{N_{Si}}\epsilon_{Ta} \text{,}
\end{equation}
where $\epsilon_{Si}$ and $\epsilon_{Ta}$ are the individual stopping powers for protons in silicon and tantalum. These were obtained using SRIM~\cite{ZIEGLER2010}. The stoichiometric ratio was determined using a yield curve analysis of the standard $622$-keV resonance, as explained in Section II.
\par
We determined a resonance strength of $\omega\gamma_{435} = (1.28 \pm 0.25) \times 10^{-4}$ eV for the 435-keV resonance (Table~\ref{tab:levels}). The primary sources of uncertainty in this measurement are the correction factor $\xi$ ($17\%$) and the ratio of stopping powers ($8\%$). A future measurement of either the resonance spin-parity or the angular correlation factors for the three primary transition would improve the uncertainty considerably.

%%%%%%%%%%%%%%%%%%%%%%%%%%%%%%%%%%%%%%%%%
\section{Thermonuclear reaction rate of $^{30}$Si($p$,$\gamma$)$^{31}$P}
\subsection{General procedure}
The reaction rate for $^{30}$Si($p$,$\gamma$)$^{31}$P depends on both resonant and non-resonant properties of the nuclear interaction. For the purposes of modeling astrophysical phenomena, an accurate description of both processes is paramount. In this section, we review the available nuclear data for the $^{30}$Si + $p$ reaction and detail the process of incorporating them into a modern reaction rate calculation.

Experimental thermonuclear rates of the $^{30}$Si(p,$\gamma$)$^{31}$P reaction were calculated using the Monte Carlo procedure presented in Longland \textit{et al.} \cite{LONGLAND2010}.
The total thermonuclear rate (in units of cm$^3$ mol$^{-1}$ s$^{-1}$) for a reaction involving two nuclei (0 and 1) in the entrance channel at a given temperature is given by
\begin{align}
    N_A \langle \sigma v \rangle = \frac{ 3.7318 \times 10^{10}}{T_9^{3/2}} \sqrt{\frac{M_0 + M_1}{M_0 M_1}} \nonumber \\
    \times \int_0^{\infty} E \sigma(E) e^{-11.605 E /T_0} dE
\end{align}
where the center-of-mass energy, $E$, is in units of MeV, the temperature, $T_9$, is in GK ($T_9 \equiv T/10^9$ K), the atomic masses, $M_i$, are in u, the cross section, $\sigma$, is in barn ($1$ b $\equiv 10^{-24}$ cm$^2$), and $N_A$ denotes Avogadro's constant.

We will briefly review the nuclear physics input necessary for the Monte Carlo rate calculations. For resonances, the cross section takes the form of the Breit-Wigner curve. If all the resonance partial widths ($\Gamma_p$, $\Gamma_\gamma$) are known, the integral in Eq.~(9) can be performed numerically. Frequently, resonances are so narrow that their cross section cannot be obtained experimentally. Instead, all that is measured is the resonance strength, $\omega\gamma$, which is proportional to the resonance integral. For each resonance, we assumed a Gaussian probability density for the resonance energy and a lognormal probability density for the resonance strength. The variance for each distribution was determined by the experimental uncertainties in the measurements. For resonances that have only been observed indirectly, i.e., using (d,n) or ($^3$He,d) reactions,
%in reactions other than $^{30}$Si($p$,$\gamma$)$^{31}$P, 
we instead estimated an upper limit, $\omega\gamma_{u.l.}$, based on the available $^{31}$P structure information (excitation energies, $E_x$, and spectroscopic factors, $C^2S$). In these cases, we assumed a Porter-Thomas probability distribution \cite{LONGLAND2010} to sample the reduced proton widths and truncated the distribution at the upper-limit value. For the recommended mean reduced width we adopted a value of $0.0003$, with an estimated uncertainty of a factor of 3 (see Figure~4 of Ref.~\cite{Pogrebnyak:2013hg}).

Finally, for the non-resonant component, the astrophysical $S$ factor is used instead of the cross section since it varies much more slowly with energy. It is expanded into a Taylor series, and the input to the rate calculation consists of the value and slope of $S(E)$ at zero bombarding energy. The probability densities of these parameters are assumed to be lognormal distributions with the uncertainties again determined by the variance.

To perform the Monte Carlo calculation we used the program \texttt{RatesMC} \cite{LONGLAND2010}, which computes a probability density function of the total rate, $N_A \langle \sigma v \rangle$, on a fixed temperature grid by sampling the relevant nuclear input. At each temperature, 20,000 samples were drawn. Based on the accumulated total rate distribution, a recommended reaction rate and rate uncertainty (assuming $68\%$ coverage) were derived. In the following we provide details on the nuclear physics input. 

%%%%%%%%%%%%%%%%%%%%%%%%%%%%%%%%%%%%%%%%%%%%%
\subsection{Observed resonances}
The $^{30}$Si($p$,$\gamma$)$^{31}$P reaction proceeds predominantly through narrow resonances. For the $E_r^{lab}=435$ keV and $E_r^{lab}=501$ keV resonances, we used the resonance strengths and energies determined in the present work (Table I).
 
Many higher-lying resonances occur in the $E_p^{lab}=0.6 - 3.0$ MeV range. A complete reference list can be found in Ref.~\cite{OUELLET2013}. We adopted resonance strength measurements from the following studies: $E_p^{lab}=671-777$ keV (Hoogenboom \cite{Hoogenboom}), $E_p^{lab}=835-983$ keV (Wolff \textit{et al} \cite{WOLFF1969}), $E_p^{lab}=1095-1516$ keV (van Rinsvelt and Smith \cite{VANRINSVELT1964}), $E_p^{lab}=1595-1830$ (van Rinsvelt and Endt \cite{VANRINSVELT1966}), $E_p^{lab}=1878-1995$ (van Rinsvelt and Endt \cite{VANRINSVELT1966}, de Neijs \textit{et al} \cite{de_Neijs_1975}), $E_p^{lab}=2010-2505$ keV (de Neijs \textit{et al} \cite{de_Neijs_1975}), $E_p^{lab}=2542-3027$ keV (Bornman \textit{et al} \cite{BORNMAN1967}). Where two references have been given, the average of the resonance measurements within the stated range was adopted. Each of these studies reported resonance strengths relative to the standard resonance at $E_r^{lab}=622$ keV. All values have therefore been scaled to the recommended value by Paine and Sargood \cite{PAINE1979} (see Section IV A).

\subsection{Unobserved resonances near the proton threshold}\label{sec:unres}
Several $^{31}$P levels near the proton threshold have been observed that may contribute to the total $^{30}$Si $+$ $p$ reaction rate (see Figure~\ref{fig:level_scheme}). We list all of these levels located below the lowest-energy resonance observed in the present work ($E_x$ $=$ $7716.0\pm0.3$~keV) in Table~II. In the following we will denote these states by their center-of-mass resonance energies, which are listed in column 2 of Table II.
\linespread{1.0}
\begin{table*}[ht!] 
\begin{threeparttable}
\caption{Properties of $^{31}$P levels near the $^{30}$Si $+$ $p$ threshold. Present values are shown in boldface.}
\center
\begin{tabular}{lrclcc}
\toprule
$E_x$ (keV)\tnote{a} & $E_r^{cm}$ (keV)\tnote{b} & $J^\pi$ \tnote{a} &   $\ell$  &   $C^2S$ \tnote{g}  &   $\omega\gamma$ (eV) \\
\colrule 
7313.7$\pm$1.6           &   18.6$\pm$1.6    &   (1/2, 3/2)$^+$                       &   0, 2      &  $\leq$0.001 &  $\leq$6.50$\times$10$^{-37}$ \tnote{h}\\
7314$\pm$4               &   18.9$\pm$4.0    &   (5/2, 7/2)$^-$                       &   3         &   0.002      &  $\approx$8.60$\times$10$^{-40}$ \tnote{h}\\
7346$\pm$6               &   50.9$\pm$6.0    &   (3/2, 5/2)$^-$                       &   1, 3      &  $\leq$1     &  $\leq$5.04$\times$10$^{-17}$ \tnote{h}\\
7356 \tnote{c}           &                   &                                        &             &              & \\
7441.4$\pm$1.0           &  146.3$\pm$1.0    &  (3/2$^+$, 5/2, 7/2, 9/2$^+$)\tnote{d} &   2, 3, 4   &  $\leq$1     &  $\leq$7.60$\times$10$^{-8}$ \tnote{h,k}\\
7442.3$\pm$0.3           &  147.2$\pm$0.3    &  11/2$^+$                              &   6         &  $\leq$1     &  $\leq$1.24$\times$10$^{-15}$ \tnote{h}\\
7466$\pm$2               &  170.9$\pm$2.0    &  (7/2, 9/2)$^-$ \tnote{f}              &   3, 5      &  $\leq$0.003 &  $\leq$1.27$\times$10$^{-10}$ \tnote{h}\\
7572 \tnote{c}           &                   &                                        &             &              &   \\
7687.2$\pm$2.0 \tnote{e} &                   &                                        &             &              &  \\
{\bf 7717.0$\pm$0.3} \tnote{i}   & {\bf 421.9$\pm$0.3} \tnote{i}   &   {\bf (3/2$^+$, 5/2$^-$)} \tnote{i}         &             &              & {\bf (1.28$\pm$0.25)$\times$10$^{-4}$} \tnote{i} \\
7736$\pm$4               & 440.9$\pm$4.0     &   (5/2, 7/2)$^-$                       &   3         & 0.02         & $\approx$3.72$\times$10$^{-4}$ \tnote{h}\\
{\bf 7781.3$\pm$0.2} \tnote{i}   & {\bf 486.2$\pm$0.2} \tnote{i}   &   3/2$^-$        &             &              & {\bf 0.188$\pm$0.014} \tnote{i} \\
{\bf 7898.0$\pm$0.3} \tnote{i}   & {\bf 602.9$\pm$0.3} \tnote{i}   &   1/2$^-$        &             &              & 1.95$\pm$0.10 \tnote{j}    \\
\botrule \\
\end{tabular}
\begin{tablenotes}
\item[a] {\footnotesize From Ref.~\cite{OUELLET2013} unless noted otherwise.}
\item[b] {\footnotesize Using $Q$ $=$ $7296.55\pm0.02$~keV \cite{Wang} and accounting for the difference in electron binding energies \cite{Iliadis:2019ch}.}
\item[c] {\footnotesize Level has only been reported in the $^{33}$S(d,$\alpha$)$^{31}$P study of Ref.~\cite{tet74}, and has been disregarded in the present work.}
\item[d] {\footnotesize Based on the $\gamma$-ray branches to a 5/2$^+$ and $7/2^+$ levels observed by Ref.~\cite{DeVoigt:1971uu}.}
\item[e] {\footnotesize Level has only been reported by Ref.~\cite{de_Neijs_1975}. It was weakly excited and the results were reported in parenthesis. We disregarded this state.}
\item[f] {\footnotesize From Ref.~\cite{ENDT1990}, based on $\gamma$-ray decay and feeding.}
\item[g] {\footnotesize Spectroscopic factors, estimated from the experiment of Ref.~\cite{VERNOTTE1990}, assuming the lowest $\ell$ value allowed.}
\item[h] {\footnotesize Assuming $\omega\gamma$ $\approx$ $0.5(2J+1)\Gamma_p$.}
\item[i] {\footnotesize From direct measurement of present work.}
\item[j] {\footnotesize From Ref.~\cite{PAINE1979}.}
\item[k] {\footnotesize Upper limit corresponds to $\ell$ $=$ 2. Values for $\ell$ $=$ 3 and 4 are given in the text.}
\end{tablenotes}
\label{tab:levels}
\end{threeparttable}
\end{table*}    
\linespread{1.5}

An unbound state at E$_x = 7466 \pm 2$ keV was discovered by Ref.~ \cite{DeVoigt:1971uu} using the $^{27}$Al($\alpha$,$\gamma$)$^{31}$P reaction, corresponding to $E_r^{cm}=170.9\pm2.0$ keV in $^{30}$Si(p,$\gamma$)$^{31}$P. This state was later confirmed by Ref. \cite{Twin} using the $^{28}$Si($\alpha$,p$\gamma$)$^{31}$P reaction. The spin-parity is restricted to (7/2, 9/2)$^-$ \cite{ENDT1990}, based on the $\gamma$-ray feeding and decay of this level. The lowest possible orbital angular momentum transfer is $\ell$ $=$ $3$. From the $^{30}$Si($^3$He,d)$^{31}$P spectrum shown in Figure~1 of Ref. \cite{VERNOTTE1990} we estimated a spectroscopic factor of $C^2S_{\ell=3}$ $\leq$ $0.003$. This value is based on the intensity of the nearby $7736$-keV peak ($C^2S = 0.02$) and assumes that $7466$-keV peak has an relative intensity of (at most) $15\%$.
Assuming $\omega\gamma$ $\approx$ $0.5(2J+1)\Gamma_p$, we find an upper limit for the resonance strength of $\omega\gamma(171)$ $\leq$ $1.27\times10^{-10}$~eV. 
\par
Ouellet \& Singh \cite{OUELLET2013} list a doublet at $7441.4\pm1.0$ keV ($J^\pi$ $=$ $3/2 - 9/2$) and $7442.3\pm0.3$ keV ($J^\pi$ $=$ $11/2^+$), corresponding to resonance energies of $E_r^{cm}=146.3\pm1.0$ keV and $147.2\pm0.3$ keV, respectively, while only one level is given by Endt \cite{ENDT1990} at $7441.2\pm0.7$ keV ($J^\pi$ $=$ $11/2^+$). The only evidence we have for concluding that these two levels are not identical is a weak (10$\pm$5\%) primary branch to a lower-lying $5/2^+$ state. Such a decay would be unlikely for a $11/2^+$ ($\ell$ $=$ $6$) state. Since the evidence for the existence of two levels is ambiguous, we will assume for the Monte Carlo sampling a 50\% chance of a contribution from a $E_r^{cm}=146.3\pm1.0$ keV resonance ($J^\pi$ $=$ $3/2 - 9/2$) that is distinct from the $147.2\pm0.3$ keV resonance (11/2$^+$). We are sampling each of the possible orbital angular momenta ($\ell$ $=$ 2, 3 or 4) for the former resonance with equal probability. Both levels are located in the $^{30}$Si($^3$He,d)$^{31}$P spectrum of Ref. \cite{VERNOTTE1990} in a region contaminated by $^{17}$F. Therefore, no more stringent estimate than $C^2S$ $\leq$ 1 can be obtained. The resulting resonance strength upper limits are $\omega\gamma(146.3)$ $\leq$ $7.60\times10^{-8}$~eV ($\ell$ $=$ 2), $\leq$ $2.76\times10^{-9}$~eV ($\ell$ $=$ 3), and $\leq$ $9.15\times10^{-11}$~eV ($\ell$ $=$ 4) for the $E_r^{cm}=146.3$ keV resonance, and $\omega\gamma(147.2)$ $\leq$ $1.24\times10^{-15}$~eV ($\ell$ $=$ 6) for the $E_r^{cm}=147.2$ keV resonance.

Ouellet \& Singh \cite{OUELLET2013} list a level at E$_x$ $=$ $7346 \pm 6$~keV ($J^\pi$ $=$ $3/2^-$, $5/2^-$), corresponding to a resonance energy of $E_r^{cm}=50.9\pm6.0$ keV. The lowest possible orbital angular momentum transfer is $\ell$ $=$ $1$. Since this level is located in the $^{30}$Si($^3$He,d)$^{31}$P spectrum of Ref. \cite{VERNOTTE1990} in a region contaminated by $^{17}$F, no more stringent estimate than $C^2S_{\ell=1}$ $\leq$ 1 can be obtained. The resulting resonance strength upper limit is $\omega\gamma(51)$ $\leq$ $5.04\times10^{-17}$~eV.

The reported \cite{OUELLET2013} doublet at $7313.7\pm1.6$ keV ($J^\pi$ $=$ $1/2^+$, $3/2^+$) and $7314.0\pm4.0$ keV ($J^\pi$ $5/2^-$, $7/2^-$) corresponds to resonance energies of $E_r^{cm}=18.6\pm1.6$ keV and $E_r^{cm}=18.9\pm4.0$ keV, respectively. The latter state was populated in the $^{30}$Si($^3$He,d)$^{31}$P study of Ref. \cite{VERNOTTE1990}, who reported a spectroscopic factor of $C^2S_{\ell=3}$ $=$ $0.002$. The resulting resonance strength is $\omega\gamma(18.9)$ $\leq$ $8.60\times10^{-40}$~eV. However, we can not exclude a contribution from the other state in the doublet, which would correspond to an s-wave resonance ($\ell$ $=$ $0$). Assuming that the entire intensity measured for this doublet in the study of Ref. \cite{VERNOTTE1990} is caused by the $7313.7\pm1.6$ keV level, and comparing this intensity to the one for the 7141 keV ($\ell$ $=$ $0$) level, we can estimate a spectroscopic factor upper limit of $C^2S_{\ell=0}$ $\leq$ 0.001. The resulting resonance strength upper limit is $\omega\gamma(18.6)$ $\leq$ $6.50\times10^{-37}$~eV.

We disregarded three levels listed in Ref.~\cite{OUELLET2013}. The level at E$_x = 7687.2 \pm 2.0$ keV is listed with a question mark and has only been reported by Ref.~\cite{de_Neijs_1975}. This state was weakly excited and the excitation energy was placed in parenthesis \cite{de_Neijs_1975}. The two levels at E$_x = 7356$~keV and 7572~keV have only been reported in the $^{33}$S(d,$\alpha$)$^{31}$P study of Ref.~\cite{tet74}. The existence of these three threshold levels is questionable at present.

\subsection{Direct capture}
The direct capture contribution has been estimated using the formalism presented in Refs.~\cite{ROLFS1973,Iliadis:2004ej}. The total direct capture cross section is given by an incoherent sum over orbital angular momenta $\ell_i$ and $\ell_f$ for all incoming and outgoing partial waves involved,
\begin{align}
    \sigma^{DC}_{total} = \sum_{\ell_i,\ell_f} C^2S(\ell_f)\sigma^{DC}_{model}(\ell_i,\ell_f)
\end{align}
where the sum runs over all bound states in $^{31}$P for which proton spectroscopic factors have been measured \cite{VERNOTTE1990}. The theoretical direct capture cross section in the energy range $E^{cm}_p$ $=$ $0.1$ $-$ $1.0$ MeV was computed using a single-particle model with a Woods-Saxon bound state potential ($r_0$ $=$ $1.25$~fm, $a$ $=$ $0.65$~fm). The total non-resonant cross section was then converted to the astrophysical S-factor, $S(E)$ $\equiv$ $\sigma_{total}^{DC}(E) E e^{2\pi\eta}$, with $\eta$ denoting the Sommerfeld parameter. The polynomial expansion of the total S-factor yields
\begin{align}
   S(E) = S(0) + S^\prime(0) E + \frac{1}{2}S^{\prime\prime}(0) E^2
\end{align}
with $S(0)$ $=$ $0.221$~MeVb, $S^\prime(0)$ $=$ $-5.52\times10^{-2}$~b, and $S^{\prime\prime}(0)$ $=$ $3.37\times10^{-2}$~b/MeV.

\subsection{Total Reaction Rate}
\begin{figure}
%0.5
\includegraphics[width=0.60\textwidth]{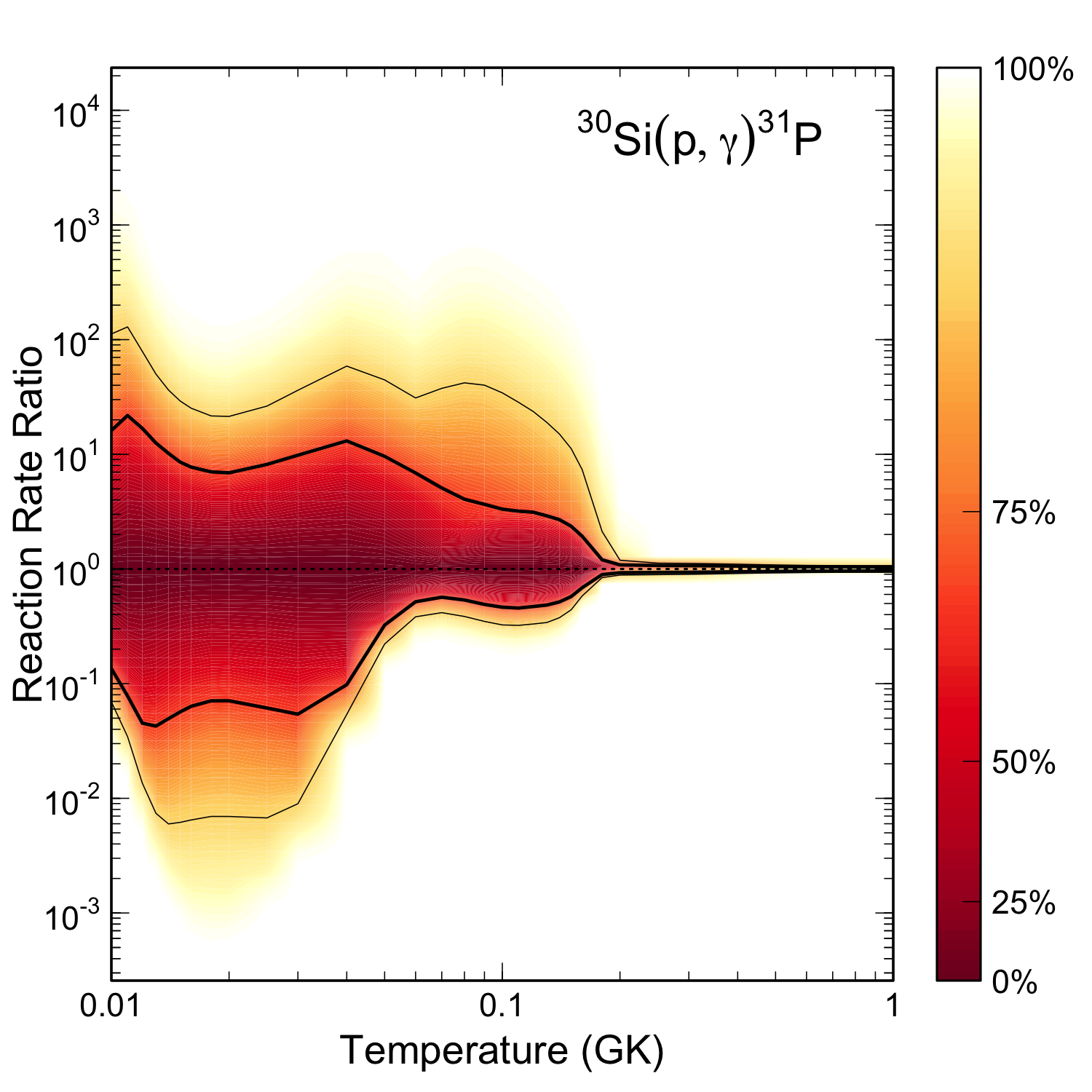}
\caption{Contour plot of the $^{30}$Si(p,$\gamma$)$^{31}$P reaction rate probability density as a function of temperature. The rate values are normalized to the recommended (median) rate. The shading indicates the coverage probability in percent. The thick and thin solid black lines indicate the high and low Monte Carlo rates for a coverage probability of $68\%$ and $95\%$, respectively.}
\label{fig:rate_uncertainty}
\end{figure}

The total $^{30}$Si(p,$\gamma$)$^{31}$P reaction rates are listed in Table~\ref{tab:rates} and are shown in Figure~\ref{fig:rate_uncertainty}. The tabulated ``low'', ``median'', and ``high'' rates refer to a coverage probability of 68\%. The tabulated rate factor uncertainty, $f.u.$, is derived from $f.u.$ $=$ $e^\sigma$, where $\sigma$ denotes the spread parameter for the lognormal approximation of the total rate probability density \cite{LONGLAND2010}. It can be seen that the reaction rate uncertainty amounts to less than 15\% at temperatures $T$ $\geq$ $0.2$~GK, but increases with decreasing temperature. For example, at $T$ $=$ $0.04$~GK and $0.1$~GK the uncertainty amounts to a factor of $\approx$9 and $\approx$4, respectively.
\linespread{0.5}
\begin{table}[ht!] 
\begin{threeparttable}
\caption{Thermonuclear reaction rates for $^{30}$Si(p,$\gamma$)$^{31}$P \tnote{a}}
\setlength{\tabcolsep}{12pt}
\center
\begin{tabular}{ccccc}
\toprule
T (GK) & Low & Median &   High  &   f.u.  \\
\colrule 
0.010 & 1.00$\times$10$^{-38}$ & 7.42$\times$10$^{-38}$ &
      1.20$\times$10$^{-36}$ & 9.983 \\ 
0.011 & 4.08$\times$10$^{-37}$ & 5.22$\times$10$^{-36}$ &
      1.14$\times$10$^{-34}$ & 13.23 \\ 
0.012 & 1.55$\times$10$^{-35}$ & 3.43$\times$10$^{-34}$ &
      5.77$\times$10$^{-33}$ & 14.64 \\ 
0.013 & 5.72$\times$10$^{-34}$ & 1.34$\times$10$^{-32}$ &
      1.68$\times$10$^{-31}$ & 14.22 \\ 
0.014 & 1.51$\times$10$^{-32}$ & 3.06$\times$10$^{-31}$ &
      3.12$\times$10$^{-30}$ & 13.18 \\ 
0.015 & 2.61$\times$10$^{-31}$ & 4.60$\times$10$^{-30}$ &
      3.93$\times$10$^{-29}$ & 12.20 \\ 
0.016 & 3.11$\times$10$^{-30}$ & 4.90$\times$10$^{-29}$ &
      3.79$\times$10$^{-28}$ & 11.52 \\ 
0.018 & 1.75$\times$10$^{-28}$ & 2.47$\times$10$^{-27}$ &
      1.74$\times$10$^{-26}$ & 10.80 \\ 
0.020 & 4.01$\times$10$^{-27}$ & 5.65$\times$10$^{-26}$ &
      3.90$\times$10$^{-25}$ & 10.75 \\ 
0.025 & 9.06$\times$10$^{-25}$ & 1.48$\times$10$^{-23}$ &
      1.21$\times$10$^{-22}$ & 11.49 \\ 
0.030 & 3.14$\times$10$^{-23}$ & 5.79$\times$10$^{-22}$ &
      5.69$\times$10$^{-21}$ & 11.89 \\ 
0.040 & 5.33$\times$10$^{-21}$ & 5.45$\times$10$^{-20}$ &
      7.13$\times$10$^{-19}$ & 9.271 \\ 
0.050 & 4.85$\times$10$^{-19}$ & 1.50$\times$10$^{-18}$ &
      1.44$\times$10$^{-17}$ & 5.432 \\ 
0.060 & 1.86$\times$10$^{-17}$ & 3.58$\times$10$^{-17}$ &
      2.46$\times$10$^{-16}$ & 4.040 \\ 
0.070 & 3.93$\times$10$^{-16}$ & 6.93$\times$10$^{-16}$ &
      3.53$\times$10$^{-15}$ & 3.976 \\ 
0.080 & 5.06$\times$10$^{-15}$ & 9.43$\times$10$^{-15}$ &
      3.83$\times$10$^{-14}$ & 4.107 \\ 
0.090 & 4.25$\times$10$^{-14}$ & 8.65$\times$10$^{-14}$ &
      3.17$\times$10$^{-13}$ & 4.140 \\ 
0.100 & 2.65$\times$10$^{-13}$ & 5.72$\times$10$^{-13}$ &
      1.90$\times$10$^{-12}$ & 4.022 \\ 
0.110 & 1.29$\times$10$^{-12}$ & 2.82$\times$10$^{-12}$ &
      9.00$\times$10$^{-12}$ & 3.832 \\ 
0.120 & 5.15$\times$10$^{-12}$ & 1.09$\times$10$^{-11}$ &
      3.43$\times$10$^{-11}$ & 3.601 \\ 
0.130 & 1.76$\times$10$^{-11}$ & 3.63$\times$10$^{-11}$ &
      1.06$\times$10$^{-10}$ & 3.347 \\ 
0.140 & 5.45$\times$10$^{-11}$ & 1.06$\times$10$^{-10}$ &
      2.84$\times$10$^{-10}$ & 3.067 \\ 
0.150 & 1.69$\times$10$^{-10}$ & 2.93$\times$10$^{-10}$ &
      6.92$\times$10$^{-10}$ & 2.715 \\ 
0.160 & 6.02$\times$10$^{-10}$ & 8.77$\times$10$^{-10}$ &
      1.69$\times$10$^{-09}$ & 2.262 \\ 
0.180 & 1.19$\times$10$^{-08}$ & 1.33$\times$10$^{-08}$ &
      1.60$\times$10$^{-08}$ & 1.472 \\ 
0.200 & 1.98$\times$10$^{-07}$ & 2.13$\times$10$^{-07}$ &
      2.31$\times$10$^{-07}$ & 1.155 \\ 
0.250 & 3.82$\times$10$^{-05}$ & 4.10$\times$10$^{-05}$ &
      4.39$\times$10$^{-05}$ & 1.073 \\ 
0.300 & 1.32$\times$10$^{-03}$ & 1.41$\times$10$^{-03}$ &
      1.51$\times$10$^{-03}$ & 1.069 \\ 
0.350 & 1.68$\times$10$^{-02}$ & 1.78$\times$10$^{-02}$ &
      1.90$\times$10$^{-02}$ & 1.064 \\ 
0.400 & 1.15$\times$10$^{-01}$ & 1.22$\times$10$^{-01}$ &
      1.29$\times$10$^{-01}$ & 1.058 \\ 
0.450 & 5.19$\times$10$^{-01}$ & 5.47$\times$10$^{-01}$ &
      5.76$\times$10$^{-01}$ & 1.053 \\ 
0.500 & 1.75$\times$10$^{+00}$ & 1.83$\times$10$^{+00}$ &
      1.92$\times$10$^{+00}$ & 1.050 \\ 
0.600 & 1.09$\times$10$^{+01}$ & 1.14$\times$10$^{+01}$ &
      1.19$\times$10$^{+01}$ & 1.045 \\ 
0.700 & 4.02$\times$10$^{+01}$ & 4.19$\times$10$^{+01}$ &
      4.37$\times$10$^{+01}$ & 1.043 \\ 
0.800 & 1.06$\times$10$^{+02}$ & 1.11$\times$10$^{+02}$ &
      1.15$\times$10$^{+02}$ & 1.042 \\ 
0.900 & 2.25$\times$10$^{+02}$ & 2.34$\times$10$^{+02}$ &
      2.44$\times$10$^{+02}$ & 1.041 \\ 
1.000 & 4.08$\times$10$^{+02}$ & 4.25$\times$10$^{+02}$ &
      4.42$\times$10$^{+02}$ & 1.041 \\ 
1.250 & 1.17$\times$10$^{+03}$ & 1.21$\times$10$^{+03}$ &
      1.26$\times$10$^{+03}$ & 1.040 \\ 
1.500 & 2.31$\times$10$^{+03}$ & 2.40$\times$10$^{+03}$ &
      2.50$\times$10$^{+03}$ & 1.040 \\ 
1.750 & 3.74$\times$10$^{+03}$ & 3.88$\times$10$^{+03}$ &
      4.04$\times$10$^{+03}$ & 1.039 \\ 
2.000 & 5.35$\times$10$^{+03}$ & 5.55$\times$10$^{+03}$ &
      5.77$\times$10$^{+03}$ & 1.039 \\ 
2.500 & 8.88$\times$10$^{+03}$ & 9.21$\times$10$^{+03}$ &
      9.56$\times$10$^{+03}$ & 1.038 \\ 
3.000 & 1.26$\times$10$^{+04}$ & 1.31$\times$10$^{+04}$ &
      1.36$\times$10$^{+04}$ & 1.038 \\ 
3.500 & 1.64$\times$10$^{+04}$ & 1.70$\times$10$^{+04}$ &
      1.77$\times$10$^{+04}$ & 1.038 \\ 
4.000 & 2.02$\times$10$^{+04}$ & 2.10$\times$10$^{+04}$ &
      2.18$\times$10$^{+04}$ & 1.038 \\ 
5.000 & 2.76$\times$10$^{+04}$ & 2.87$\times$10$^{+04}$ &
      2.98$\times$10$^{+04}$ & 1.039 \\ 
6.000 & {\it 3.64$\times$10$^{+04}$} & {\it 3.89$\times$10$^{+04}$} & {\it 4.16$\times$10$^{+04}$} &   1.040 \\ 
7.000 & {\it 4.74$\times$10$^{+04}$} & {\it 5.06$\times$10$^{+04}$} & {\it 5.41$\times$10$^{+04}$} &   1.041 \\ 
8.000 & {\it 5.80$\times$10$^{+04}$} & {\it 6.20$\times$10$^{+04}$} & {\it 6.63$\times$10$^{+04}$} &   1.042 \\ 
9.000 & {\it 6.87$\times$10$^{+04}$} & {\it 7.34$\times$10$^{+04}$} & {\it 7.84$\times$10$^{+04}$} &   1.044 \\ 
10.00 & {\it 8.13$\times$10$^{+04}$} & {\it 8.69$\times$10$^{+04}$} & {\it 9.29$\times$10$^{+04}$} &   1.045 \\ 
\botrule \\
\end{tabular}
\begin{tablenotes}
\item[a] {\footnotesize In units of cm$^3$mol$^{-1}$s$^{-1}$. Columns 2, 3, and 4 list the 16th, 50th, and 84th percentiles of the total rate probability density at given temperatures. Rates for $T$ $\geq$ $6$~GK have been adopted from Ref.~\cite{ILIADIS2010}; they have not been obtained from the Monte Carlo sampling procedure, but account for the contributions of higher-lying resonances using the Hauser-Feshbach model.}
\end{tablenotes}
\label{tab:rates}
\end{threeparttable}
\end{table}    
\linespread{1.5}

\begin{figure}[ht!]
%0.5
\includegraphics[width=0.60\textwidth]{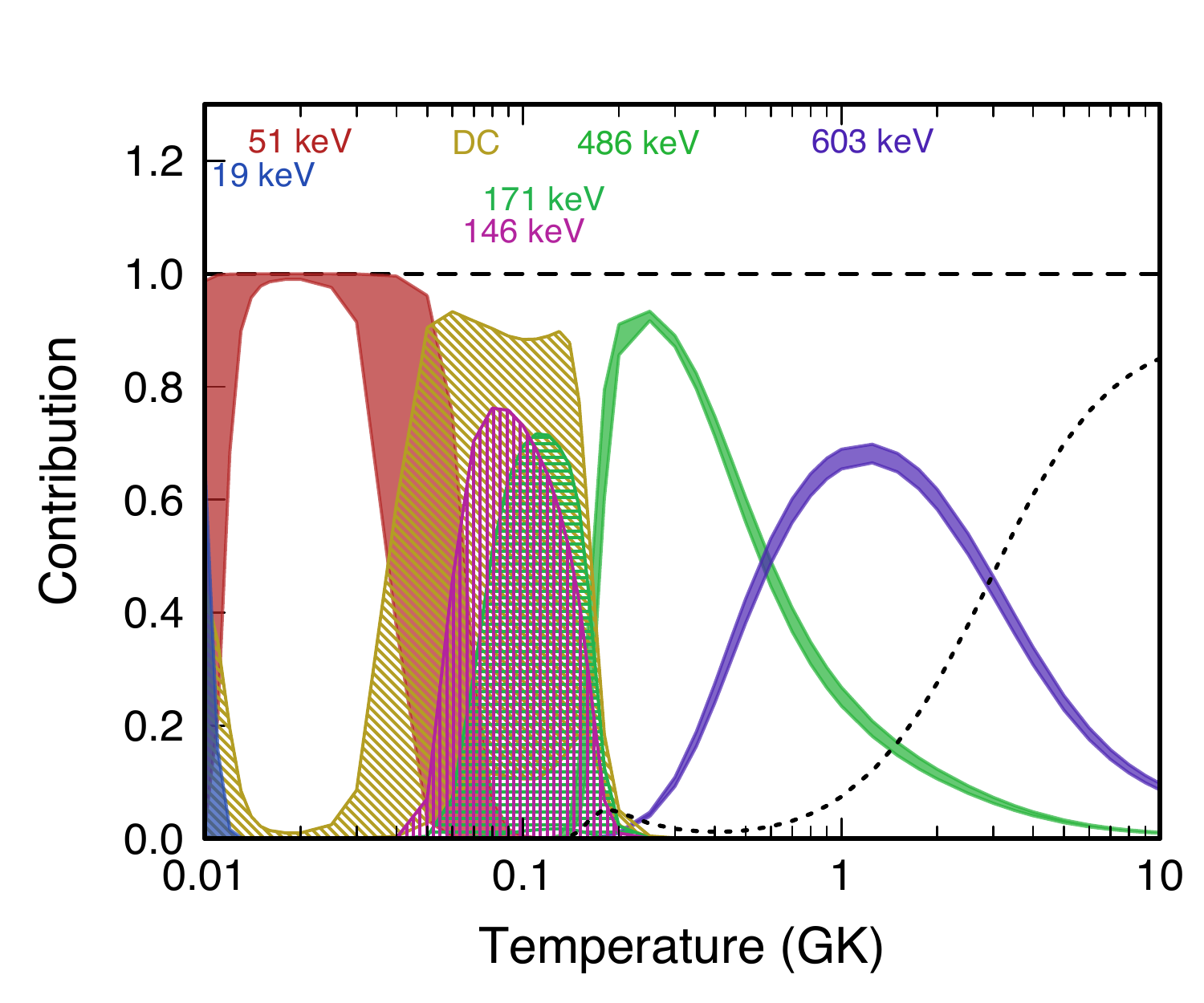}
\caption{The fractional contributions made by $^{30}$Si(p,$\gamma$)$^{31}$P resonances and direct capture (labeled ``DC'') toward the total reaction rate. The contribution ranges are shown as colored bands that correspond to their label above. The thickness of each band represents the uncertainty of the contribution. The dotted black line shows the contribution of resonances with energies larger than $648$ keV.}
\label{fig:rate_contributions}
\end{figure}

Figure~\ref{fig:rate_contributions} shows the fractional contributions of individual resonances to the total $^{30}$Si(p,$\gamma$)$^{31}$P reaction rate. Resonances with energies above $E_r^{cm}$ $=$ $648$ keV dominate the rate for temperatures $T$ $\geq$ $3.0$~GK. The 603~keV resonance, the highest-energy resonance measured in the present work, determines the rates in the range of 0.6~GK to 3~GK. The temperature region from 0.15~GK to 0.6~GK is dominated by the 486 keV resonance, also measured in this work. Between 0.04~GK and 0.15~GK, the resonances at 51~keV, 146~keV, 171~keV, and the direct capture process, contribute significantly to the total rates. Notice that only upper limit contributions could be established for these three resonances. The resonances at 147~keV, and 440~keV, as well as the 422 keV resonance measured in the present work, provide only insignificant contributions. Below $T$ $=$ $0.04$~GK, the 51~keV resonance is the most important contributor to the total rate.

\begin{figure}[h!]
%0.5
\includegraphics[width=0.6\textwidth]{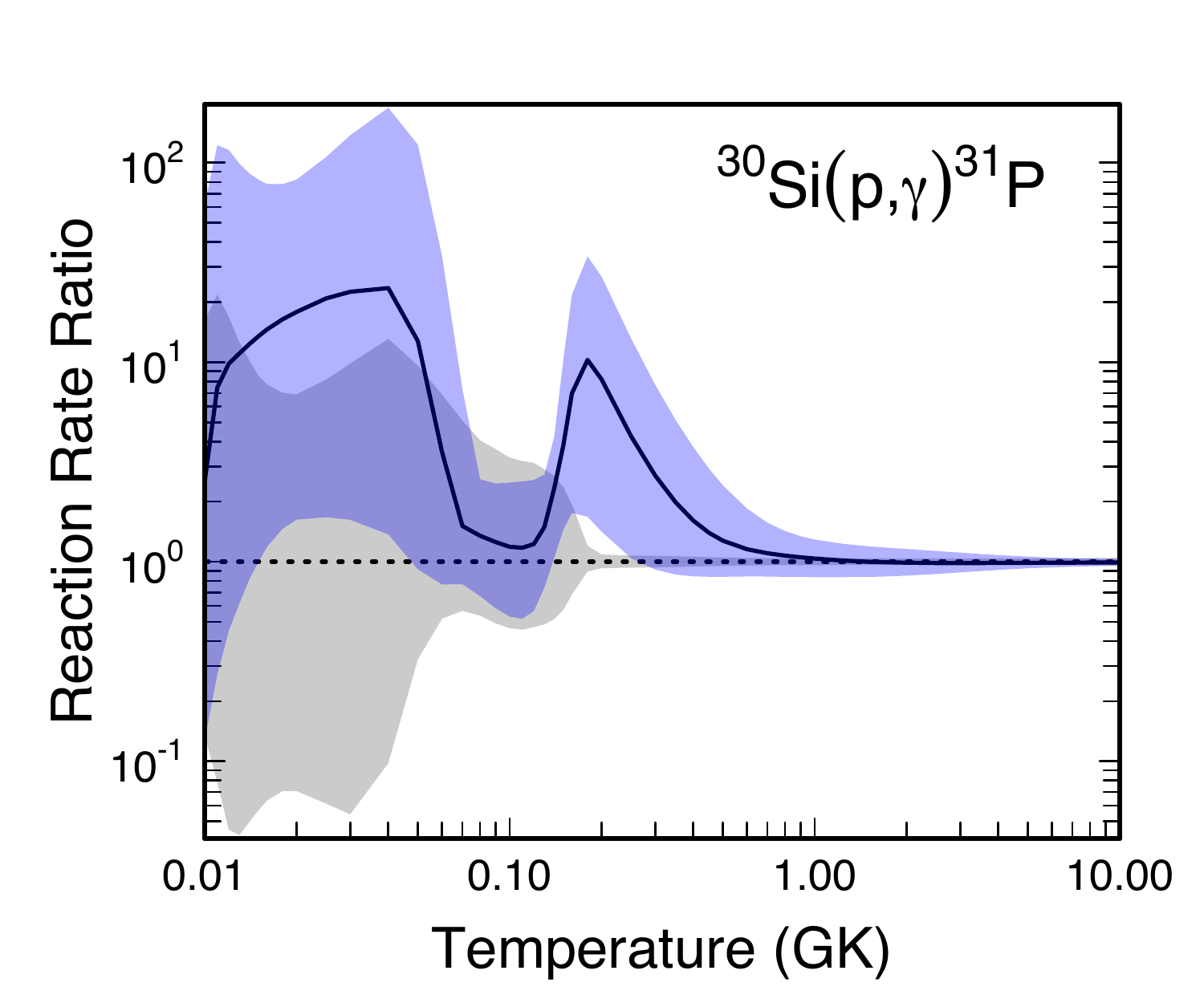}
\caption{Reaction rates from the present work (gray) and the evaluation of Iliadis \textit{et al.}~\cite{ILIADIS2010} (blue), normalized to the present recommended rates. The shaded areas correspond to $68\%$ coverage probabilities. The black solid line shows the ratio of the two recommended rates. Notice that the gray shaded area is the same as the area between the thick solid lines in Figure~\ref{fig:rate_uncertainty}.}
\label{fig:rate_compare}
\end{figure}

It is instructive to compare the present rates with those from the 2010 evaluation \cite{ILIADIS2010}. The gray and blue shaded areas in Figure~\ref{fig:rate_compare} show the 68\% coverage probabilities of the present and previous rate probability density, respectively. All rates are normalized to the present recommended (median) rate listed in column 3 of Table~\ref{tab:rates}. The previous rates are higher than the present results by about an order of magnitude near $T$ $=$ $0.2$~GK, which is a direct consequence of our measurement of the previously unobserved 422 keV resonance. We have reduced the contribution of this resonance to an insignificant level. From the respective widths of the gray and blue bands, it can also be seen that the present measurement has drastically reduced the total rate uncertainty near $0.2$~GK from a factor of $3.5$ to about 15\%.

At lower temperatures, the situation is more complex. Near $0.1$~GK it can be seen that the total rate uncertainty has slightly increased compared to the 2010 evaluation \cite{ILIADIS2010}. The reason is that only a single threshold state at $E_x$ $=$ $7441.2$~keV (11/2$^+$) was reported by Endt \cite{ENDT1990}, whereas in the present work we take a doublet at $7441.4$~keV ($3/2$ $-$ $9/2$) and $7442.3$~keV ($J^\pi$ $=$ $11/2^+$) into account \cite{OUELLET2013}, as discussed in Section~\ref{sec:unres} and shown in Table~\ref{tab:levels}. The first level is associated with a d-wave ($\ell$ $=$ $2$) resonance at $E_r$ $=$ $146.3$~keV, which would likely contribute significantly more to the total rate compared to the second level corresponding to an $\ell$ $=$ $6$ resonance at 147.2~keV. 

\section{Summary}
The $^{30}$Si(p,$\gamma$)$^{31}$P reaction was previously identified as being critical to our understanding of the abundance anomalies observed in some globular clusters. In the present work, we obtained new strength measurements for two resonances,  at $E_r^{lab} = 434.6 \pm 0.3$ keV and $E_r^{lab} = 501.1 \pm 0.2$ keV. For the former, we reported the first resonance strength based on a direct measurement, $\omega\gamma_{435} = (1.28\pm0.25) \times 10^{-4}$ eV. Using the $\gamma$-ray decay signature, the spin-parity of this state was restricted to $J^{\pi} = (3/2^+, 5/2^-)$. For the latter, we obtained a resonance strength of $\omega\gamma_{501}=(1.88\pm0.14)\times10^{-1}$ eV, which has a smaller uncertainty than previous results.

These strength measurements will help us to better understand the resonant component to the $^{30}$Si(p,$\gamma$)$^{31}$P reaction rate. The $E_{r}^{lab}$ $=$ 501~keV resonance, which was previously thought to play a minor role, is now understood to be the dominant resonance at stellar temperatures between 0.15 GK to 0.6 GK. Conversely, the contribution of the $E_{r}^{lab}$ $=$ 435-keV resonance is found to be insignificant, contrary to previous results.

New thermonuclear rates for $^{30}$Si(p,$\gamma$)$^{31}$P were presented. The reaction rate at 0.2 GK has been reduced by a factor of 10. Furthermore, the rate uncertainty has been reduced at temperatures near 0.2 GK from a factor of 3.5 to about $15\%$. The implications of our new rate for the hydrogen burning in globular cluster stars will be explored in forthcoming work.

\begin{acknowledgments}
We would like to thank Sean Hunt for his help with preparing the targets. This work was supported in part by the U.S. DOE under contracts DE-FG02-97ER41041 (UNC) and DE-FG02-97ER41033 (TUNL).  
\end{acknowledgments}

\bibliography{main}

\end{document}